\documentclass[preprint2]{aastex} 

\shorttitle{An Unusual Burst from SGR~1900+14}
\shortauthors{Ibrahim, Alaa I. et al.}

\begin{document}

\title{An Unusual Burst from Soft Gamma Repeater SGR~1900+14:\\
Comparisons with Giant Flares and Implications for the Magnetar Model}

\author{
Alaa I. Ibrahim\altaffilmark{1,2},
Tod E. Strohmayer\altaffilmark{1},
Peter M. Woods\altaffilmark{3,4},
Chryssa Kouveliotou\altaffilmark{3,4},
Christopher Thompson\altaffilmark{5},
Robert C. Duncan\altaffilmark{6},
Stefan Dieters\altaffilmark{7},\\
Jan van Paradijs\altaffilmark{7,8},
Mark Finger\altaffilmark{3,4}}

\altaffiltext{1}{NASA Goddard Space Flight Center, Laboratory for High Energy Astrophysics,
 Greenbelt, MD 20771; Alaa@milkyway.gsfc.nasa.gov} 
\altaffiltext{2}{Department of Physics, George Washington University} 
\altaffiltext{3}{NASA Marshall Space Flight Center, SD--50, Huntsville, AL 35812} 
\altaffiltext{4}{Universities Space Research Association} 
\altaffiltext{5}{Department of Physics and Astronomy, University of North Carolina, Chapel Hill, NC 27599}
\altaffiltext{6}{Department of Astronomy, University of Texas, Austin, TX 78712}
\altaffiltext{7}{Department of Physics, University of Alabama in Hunstville, Hunstville, AL 35812} 
\altaffiltext{8}{Astronomical Institute "Anton Pannekoek," University of Amsterdam, Kruislaan 4003, 1098 SJ 
Amsterdam, The Netherland}

\slugcomment{Submitted to the Astrophysical Journal} 

\begin{abstract}

The Soft Gamma-ray Repeater SGR~1900+14 entered a remarkable phase of activity during the
summer of 1998. This activity peaked on August 27, 1998 when a giant periodic $\gamma$-ray flare resembling 
the famous March 5, 1979 event from SGR~0526--66 was recorded. Two days later (August 29), a strong, bright 
burst was detected simultaneously with the Rossi X-ray Timing Explorer (RXTE) and the Burst and Transient 
Source Experiment (BATSE). This event reveals several similarities to the giant flares of August 27 and March 5 
and shows a number of unique features not previously seen in SGR bursts.
Unlike typically short SGR bursts (duration $\sim 0.1$ s), this event features a 3.5 s burst peak that was 
preceded by an extended ($\sim 1$ s) complex precursor, and followed by a long ($\sim 10^{3}$ s) periodic 
tail modulated at the 5.16 s stellar rotation period. The tail also shows several short recurrent bursts.
Spectral analysis shows a striking distinction between the spectral behavior of the precursor, main peak and 
long tail. While the spectrum during the peak is uniform, a significant hard-to-soft spectral evolution is 
detected in both the precursor and tail emissions. 
Temporal behavior shows a sharp rise ($\sim$ 9.8 ms) at the event onset and a rapid cutoff ($\sim$ 17 ms) at 
the end of the burst peak. The tail pulsations show a simple pulse profile consisting of one 5.16 s peak that did 
not evolve with time. The spectral and temporal signatures of this event imply that the precursor, main peak, 
and extended tail are produced by different physical mechanisms.
We discuss these features and their implications in the context of the magnetar model. The bright 3.5 s component 
is consistent with a very hot ($T \sim 1$ MeV) trapped fireball, and the precursor with magnetospheric emission in 
which the radiating particles are heated more continuously. Less than a percent of the fireball energy will be
conducted into the exposed surface of the neutron star, thereby dissociating heavy elements and even helium, 
and inducing rapid transformations between neutrons and protons. The extended `afterglow' tail of the August 29 burst 
is consistent with a cooling hotspot of a very small area ($\sim 1.3$ km$^2$), and indicates that the energy release 
in an SGR burst is strongly localized. The energetics of the August 29 event, and its close proximity to the August 27 
flare, suggest that it is an `aftershock' of the preceding giant flare.

\end{abstract}



\clearpage

\clearpage

\section{Introduction}

Soft Gamma Repeaters (SGRs) are high-energy transient astrophysical sources that have been
identified as young neutron stars associated with persistent X-ray counterparts and supernova
remnants (SNRs). There are only four known SGRs; three within our galaxy (SGR~1900+14, SGR~1806--20, and 
SGR~1627--41) and one in the Large Magellanic Cloud (SGR~0526--66). A fifth candidate, SGR~1801--23, has not been
well-localized yet (\markcite{C2000}Cline et al.~2000). These sources are characterized by their recurrent
emission of brief ($\sim0.1$ s), intense ($\sim10^{3}-10^{4}\ L_{Edd}$) bursts with non- or modestly
varying soft $\gamma$-ray spectra (\markcite{F94}Fenimore et al.~1994; \markcite{SI97}Strohmayer \& Ibrahim
1997; \markcite{W99}Woods et al.~1999b). Burst emission from these sources tends to be concentrated into
short periods (weeks to months) of intense activity separated by relatively long periods (years) of
quiescence (\markcite{K95}Kouveliotou et al.~1995).

Recently, it was discovered that two SGRs (1806--20 and 1900+14) are also persistent X-ray
pulsars which spin down at a rapid rate of $\sim$ 10$^{-10}$ s s$^{-1}$ (\markcite{K98}Kouveliotou et al.~1998, 
1999; \markcite{H99c}Hurley et al.~1999c) . Kouveliotou et al.~have attributed this spin-down to 
magnetic braking and the corresponding magnetic fields are found to be greater than $10^{14}$ G. This provides 
strong evidence that SGRs are highly magnetized neutron stars, i.e. {\it magnetars}, an idea first proposed by 
\markcite{DT92}Duncan \& Thompson (1992). In a magnetar, the magnetic field is the dominant source 
of free energy, greater even than the rotational energy of the star. It is this energy source which 
is likely tapped to generate the recurrent bursts of $\gamma$-rays. When magnetic stresses build up sufficiently 
to crack a patch of the neutron star crust, the resulting ``crustquake'' ejects hot plasma particles (fireball) into the 
magnetosphere, which result in an SGR burst (\markcite{TD95}Thompson \& Duncan 1995).

After a long period of quiescence lasting more than five years (\markcite{K93}Kouveliotou et al.
1993), SGR~1900+14 entered a phase of extreme burst activity starting in May 1998
(\markcite{H99c}Hurley et al.~1999b). This period of enhanced burst activity has now ended, but 
not before hundreds of bursts were emitted by the source. The pinnacle of this active period was 
reached on August 27, 1998 when a giant $\gamma$-ray flare was detected with multiple spacecraft 
(\markcite{C98}Cline et al.~1998; \markcite{H99a}Hurley et al.~1999a; \markcite{Fer99}Feroci et al.~
1999; \markcite{Maz99a}Mazets et al.~1999a) and by its affect on the earth's ionosphere (\markcite{In99}Inan
et al.~1999). This flare closely resembles the famous March 5 1979 event from SGR~0526--66 
(\markcite{Maz79}Mazets et al.~1979) in that it reached a much higher peak luminosity than typical SGR bursts 
($10^{43}-10^{44}$ ergs s$^{-1}$), released a large amount of energy ($\gtrsim 10^{44}$ ergs), and persisted 
for a long time ($\sim300$ s) during which the $\gamma$-ray intensity was clearly modulated by the stellar 
rotation period. In the magnetar model, giant flares are triggered by subsurface motions in which the internal 
magnetic field rearranges itself into a lower energy state. This instability induces reconnection 
and large-amplitude wave motions in the magnetosphere, which rapidly dissipate into a hot fireball 
that gives rise to an observable $\gamma$-ray flare.

The August 27 giant flare can be separated into roughly three distinct regions: a soft, short 
precursor, a hard, bright initial pulse, and $\sim5$ minute long oscillatory tail. The initial 
pulse has a much harder spectrum which is qualitatively similar to what was seen in the March 5 
event (\markcite{H99a}Hurley et al.~1999a). During the course of the oscillating tail, the 
spectrum varied only modestly (from hard to soft), but the pulse profile changed dramatically from a 
complex four-pronged profile to a more nearly sinusoidal profile near the end of the burst 
(\markcite{Fer99}Feroci et al.~1999, \markcite{Maz99a}Mazets et al.~1999a, Feroci et al.~2000). 
Furthermore, the pulse profile of the persistent emission from this pulsar has changed accordingly. 
In all observations prior to 27 August 1998, the profile ($2-10$ keV) is fairly complex, but all 
observations to date following this flare show a nearly sinusoidal profile over the same energy 
range (\markcite{K98}Kouveliotou et al.~1999, \markcite{W99}Woods et al.~1999a).

Following the August 27 flare, a series of public observations of SGR~1900+14 were initiated with the 
Proportional Counter Array (PCA) aboard the {\it Rossi X-ray Timing Explorer} (RXTE). On 29 August 1998, 
a strong, bright burst was observed simultaneously with RXTE (Fig.~1a) 
and the Burst and Transient Source Experiment (BATSE) (Fig.~1b). We note that this event occurred only 47
hours and 55 minutes after the onset of the August 27 flare, and that it was also seen by the Ulysses 
and BeppoSAX Gamma-ray Burst Monitor (\markcite{Hur2000}Hurley et al.~2000). This event shared some 
similarities with both short SGR outbursts, and the March 5 and August 27 giant flares. Its 
bright peak component appears to have been an unusually long version of the short SGR bursts,
as indicated by its spectral hardness and weak spectral evolution. Large amplitude pulsations were 
present in a tail extending for hundreds of seconds which, in spite of not being of the same nature
as the pulsating phases of the giant flares, appears to represent a new phenomenon in the SGR sources. 
This tail was much fainter and showed stronger spectral softening than the pulsating phases of the 
two giant flares. In addition, its flux did not terminate sharply but appears instead to have 
smoothly merged with the persistent emission of SGR~1900+14. 

In this paper we present a detailed study of the spectral and temporal behavior of the August 29 event 
and compare it with the two giant flares of August 27 and March 5.


\section{BATSE Observations}

On August 29, 1998, BATSE triggered at 10:16:32.5 UT on a bright burst whose location was consistent 
with SGR~1900+14. This association was later confirmed through construction of an IPN annulus that 
included the SGR (\markcite{H99d}Hurley et al.~1999d). This event has a very smooth temporal profile 
lasting $\sim$ 3.5 s with a very abrupt beginning and end to the burst (Fig.~1b). Furthermore, there 
is a weak precursor which starts $\sim$ 1 s before the onset of the main burst. No significant emission 
from the SGR is seen in the BATSE data following this burst for more than 8 hours. Only two out of 
$\sim$ 200 bursts from SGR~1900+14 observed with BATSE in that phase of activity (\markcite{W99b}Woods et al.~
1999b) have had such smooth profiles and relatively long durations: this event and a burst 
recorded on 1998 October 28. Unfortunately, the second burst was not observed with RXTE/PCA.

The precursor was too weak in the BATSE energy band to reconstruct a meaningful spectrum, but the burst
itself yielded multiple spectra. The binning of the data on-board provided 10 time bins during which
high-quality spectra (256 energy channels) were available covering 25 keV to 4 MeV. As is traditionally done
for SGR burst spectra (\markcite{F94}Fenimore et al.~1994), we fit an optically thin thermal 
bremsstrahlung (OTTB, model bremss in XSPEC) model to the data. Based upon the low reduced 
$\chi^2_{\nu}$ values, we find this function well represented these data. 
The bremsstrahlung temperature, $kT$, did not change significantly through the burst interval, 
despite a factor $\sim$ 4 change in flux. For the time-integrated burst spectrum, we find $kT$ = 
20.6 $\pm$ 0.3 keV and the fluence ($>$ 25 keV) = (1.88 $\pm$ 0.08) $\times$ 10$^{-5}$ ergs 
cm$^{-2}$. The peak flux on the 0.064 s timescale is F$_{\rm peak}$ = (1.31 $\pm$ 0.08) $\times$ 
10$^{-5}$ ergs cm$^{-2}$ s$^{-1}$. For an assumed distance of 7 kpc for SGR~1900+14 (\markcite{Vas}Vasisht et 
al.~1994), we find a peak luminosity L$_{\rm peak}$ $\approx$ 8 $\times$ 10$^{40}$ ergs s$^{-1}$ 
and a total burst energy E$_{\rm burst}$ $\approx$ 1 $\times$ 10$^{41}$ ergs.

When compared to other bursts from SGR~1900+14 observed with BATSE, this event is one of the most 
energetic due to its long duration, but it does not have an exceptionally large peak flux. There are 
at least three events with larger peak fluxes that have more typical SGR burst durations of less 
than one second. We conclude this peculiar feature of a smooth temporal profile with a sharp 
beginning and end is not entirely determined by the peak intensity of the burst.


\section{RXTE/PCA Observations}

A series of public Target of Opportunity (TOO) observations of SGR~1900+14 with RXTE began on 29 August
1998. Shortly into the first day of observations, the bright burst which triggered BATSE (described above)
was recorded with the PCA. Due to a much lower background than BATSE, its large area, high
sensitivity, and a more optimal SGR emission bandpass ($2-60$ keV), this event was seen in much
greater detail with the PCA. Unfortunately, the event was so bright that the detector was
saturated during the majority of the burst peak. However, the enhanced sensitivity
allowed us to observe the precursor and the extended tail emission following the peak of the burst.

As seen in Fig.~1c, the precursor which is very weak in the BATSE light curve is easily
visible here and shows significant substructure. Fig.~1a gives an expanded view of the PCA light 
curve that reveals an extended tail of emission which lasts more than 1000 s beyond the peak of the 
burst. The tail is clearly modulated at the stellar rotation period of 5.16 s. Superposed on this 
tail are several smaller bursts similar to typical SGR burst 
emissions.

We have performed spectral fits to the precursor, burst rise and fall intervals, and tail emission. 
Several models produced acceptable fits, but the best fits were obtained with an OTTB model and a 
combination of power law (PL) plus blackbody radiation (BB), each modified by photoelectric absorption. 
As background, we chose $\sim 1000$ s of pre-burst data, with no other burst emission, but still containing 
persistent emission from SGR~1900+14. Our resulting spectral fits are therefore of the burst emission only. 
We have also tracked the spin of the pulsar before and after this burst to search for a discrete change 
(i.e. glitch) at the time of the burst.

We begin this section by addressing the deadtime and pile-up effects on the data then present the spectral 
analysis results for the different components of the event.


\subsection{Deadtime and Pile-up Issues}

The importance of deadtime effects in the PCA during this burst are illustrated in Fig.~2 which 
shows the good X-ray event rate, $r_G$ (solid), and the so-called remaining counts rate, $r_{\rm rem}$ 
(dashed), during the burst. Both of these rates are tabulated every 1/8 s in the Standard1 data mode of 
the PCA Experiment Data System (EDS). The remaining counts rate includes all events which triggered more 
than one anode in the PCA and at modest source counting rates gives a measure of the particle background 
rate. During the main burst, the remaining counts rate is dominated by multiple anode events due to the 
high X-ray flux from the source. During the precursor, the deadtime fraction is dominated by the good 
counting rate. This is the simplest regime to treat deadtime effects in the PCA (see \markcite{J98}Jahoda 
et al.~1998; and \markcite{S97}Strohmayer et al.~1997). In the precursor the event rate briefly reaches 
about 90,000 s$^{-1}$ in the PCA, but the average rate is only about 15,000 s$^{-1}$. At the peak rate of 
the precursor deadtime effects approach $\approx 30 \%$, but on average the deadtime is $ < 10 \%$ 
(see for example \markcite{S97}Strohmayer et al.~1997).

Note that the remaining counts rate dominates the good event rate during the main peak of the
burst. The deadtime in this regime is extreme and we do not use data from this interval to
generate spectra. Rather, we only investigate spectra for intervals in which
the peak good rate is less than about 90,000 s$^{-1}$ {\it and} the remaining counts rate is less 
than the good event rate. This criterion amounts to a rejection of any data during intervals when 
the deadtime as estimated using the procedure outlined by \markcite{J98}Jahoda et al.~(1998) is more 
than about 30 \%. Using the spectrum and peak flux observed by BATSE during the main peak of the 
burst we can estimate a peak incident contrite on the PCA of $\approx 1\times 10^7 $ s$^{-1}$, 
which is about 100 times the maximum throughput of the instrument, further confirming a huge 
deadtime problem during the peak of the burst.

An instrumental effect which could conceivably alter the inferred spectral temperature at high
counting rates, and thus any conclusions regarding spectral evolution, is pulse pile-up in the PCA. 
We have used the pulse pile-up correction procedure recently outlined by \markcite{TK98}Tomsick \& 
Kaaret (1998) to investigate the possible extent of pile-up effects on our spectral analysis. To 
estimate the magnitude of any pile-up effects we computed model counts rate spectra by folding OTTB 
models, including photoelectric absorption, with parameters similar to those inferred from the BATSE 
and PCA spectral analysis, through the PCA response function and then applied the pile-up model in 
order to simulate the effects of real pile-up in the detectors. We then generated Poisson 
realizations of the model counts rate spectra using both the piled-up and unpiled spectra. We fit 
these simulated spectra to the same model and compared the derived temperatures and absorbing 
columns. As an example we describe results for one set of simulations which were representative of 
the magnitude of the effects expected at the highest observed counts rates during the precursor. For 
an input OTTB spectrum with $kT = 30$ keV and $n_H = 3.75 \times 10^{22}$ we found that the pile-up 
effect {\it increases} the inferred temperature by $\approx 10 \%$ and decreases the inferred column 
density by about 2 \%. 


\subsection{Precursor}

The precursor to this burst has a rapid onset, rising to a peak in 9.8 ms and a relatively long 
duration of $\sim 1$ s. An exploded view of the time history of the precursor is shown in Fig.~
3a. There is no clear distinction between the end of the precursor and the start of the main burst 
emission in the PCA data since the counts rate is not observed to return to background.

We fit the OTTB model to several intervals during the precursor that met our deadtime constraints. 
In order to acquire spectra with similar statistical quality, we divided the precursor into 5 
intervals each of which contained the same number of source counts. We find significant spectral 
evolution throughout the precursor (see Table 1). The results are illustrated in Fig.~3 which shows (a) 
the precursor light curve, (b) the hardness ratio ($11-50$ keV)/($2-10$ keV), (c) the OTTB temperature, and (d) 
the inferred column density. The temperature drops significantly from 28.7 $\pm$ 4.3 keV to 11.8 
$\pm$ 1.0 keV before recovering somewhat to 20 keV just before the burst rise. To quantify the spectral 
deviations we fit a constant temperature to the 5 intervals and confirm that a constant OTTB temperature 
during the precursor is rejected at $\approx 2\times 10^{-4}$ significance. We find no evidence for a 
correlation between intensity and hardness. However we note that interval 4 which contains the highest 
counts rate during the precursor also shows the softest spectrum with the lowest $kT$.

We expect that the shift in $kT$ between the piled-up and unpiled spectra, of order $10 \%$, 
represents an upper limit to the shift which could be introduced by pile up effects during the 
precursor. We are confident of this conclusion for several reasons: 1) We estimated the effect by 
normalizing the counts rate to its peak value measured during the precursor, which exceeded the 
average counts rate in all intervals $1-5$; 2) Within each interval, the {\it variation} in the counts 
rate was smaller than the full range represented in Fig.~3a; 3) The softest intervals near the 
beginning of interval 4 correspond to the {\it highest} intensity observed in the precursor. 
Pile-up would act to harden the spectrum, indicating that the intrinsic incident spectrum would be 
even softer than inferred. For these reasons, we conclude that the observed changes in $kT$ 
cannot be entirely due to instrumental effects and that they represent real physical changes within 
the source.


\subsection{Burst Emission}

Following the precursor, the main burst emission rises sharply (in $\sim$ 20 ms) to PCA 
saturation and persists for 3.5 s before decaying abruptly (see Fig.~1). We extracted PCA spectra 
for only the rising and falling edges of the main burst emission on account of the high deadtime 
during PCA saturation. We selected intervals during which the counting rates were 
$<$ 90,000 counts s$^{-1}$, using the same criterion as described above for the precursor so as not to 
be dominated by deadtime effects. We fit the data with the OTTB model and found that the spectra of 
the rising and falling edges were consistent, with $kT = 20.0 \pm 2.3$ keV, $n_H = 3.3 \pm 1.9 
\times 10^{22}$ cm$^{-2}$ for the rising edge and $kT = 20.9 \pm 2.1$ keV, $n_H = 5.48 \pm 1.8 
\times 10^{22}$ cm$^{-2}$ for the falling edge. This, combined with the BATSE result suggests a 
non-varying spectrum with an average $kT$ of about 20 keV for the main
portion of the burst, in contrast with the precursor spectrum which shows significant changes in 
temperature.


\subsection{Pulsating Tail Emission}

A clear transition from the main burst peak to an extended tail is seen in Fig.~1a and 1d. 
The main burst emission shows an abrupt cutoff on a timescale of $\sim$17 ms, beyond which the 
X-ray flux remained well above the pre-burst level. This residual emission decayed slowly over about 
1000 s, forming the extended ``afterglow" tail. Our first step in analyzing the tail was to remove, 
by visual inspection, the small overlying bursts from the light curve. We then broke up the tail into 8 
intervals, increasing the integration time in successive intervals so as to keep the total number of 
counts (after subtraction of the pre-burst emission as background) consistent among the different 
intervals. 

The intervals used to investigate the tail spectrum are shown in Fig.~4a, the 5.16 s pulsations are 
also evident here. We first fit the spectrum of the tail to the OTTB model (see Table 1). We found 
that the OTTB model gave reasonable fits to some of the intervals but not all, however it does give a 
simple characterization of the spectral continuum. To investigate spectral evolution during the tail 
we plotted the derived OTTB temperature and the inferred column density for the 8 intervals as shown 
in Fig.~4b and 4c. We found that a combination of a power law plus a black body spectrum (PL+BB) 
generally produced a better fit to the tail intervals. The parameters of this model through the tail 
are shown in table 2. The fitted evolution of the black body component is consistent with a hot spot 
of constant radius $\sim 0.6$ km. 

The temporal behavior of the flux of the X-ray tail can be characterized by a power-law decay with an 
exponent $\alpha = 0.8\pm0.1$. With the smaller bursts removed, background subtracted, and correcting for 
RXTE offset pointing of $0^{\circ}\!\!.441$, the total fluence and energy released in the tail are 
7.5 $\times 10^{-8}$ ergs cm$^{-2}$ and 4.4 $\times 10^{38}$ ergs respectively (assuming a distance of 7 kpc).


\subsection{Pulse Phase Spectral Analysis}

The pulsations throughout the tail have a simple pulse profile as shown in Fig.~5a. The ephemeris 
of the pulsar during this time period was determined precisely elsewhere (\markcite{W99a}Woods et 
al.~1999a). Using a phase folding technique, we fit the data around this burst with finer sampling 
intervals than before in order to search for any discontinuous changes in phase at the time of this 
event. We do not find any significant deviation in frequency which places a limit on a glitch 
associated with this burst of $|\Delta P/ P| < 5.4\times 10^{-5}$.

The bursts during the tail were again removed, and using the ephemeris derived earlier, we folded 
the tail over different energy bands. The phase folded profile of the tail emission is nearly 
sinusoidal following this burst which is similar to what is seen in the persistent emission 
following the August 27 flare (\markcite{K99b}Kouveliotou et al.~1999; \markcite{Mur99}Murakami et 
al.~1999; \markcite{W99a}Woods et al.~1999). Comparison of the light curves in different energy 
bands shows there is spectral evolution over the phase of the pulsar during the tail (Fig.~5b and 5c). 
The data were binned in phase according to our model and fit for five different phase intervals. We find 
the maximum of the pulsed emission is slightly hotter than elsewhere in phase (Fig.~5b).


\section{Comparisons with the August 27 and March 5 Giant Flares}

The August 29 event and the preceding giant flare on August 27, 1998 are -- together with the March 5, 1979
giant flare from SGR~0526--66 -- the only SGR outbursts whose observed emissions lasted long enough to show a
clear modulation at the stellar rotation period. Table 3 compares the properties of these three remarkable
events.

The August 29 event released less energy than either giant flare by a factor $\lesssim 10^{-3}$ if the
emission was isotropic (as assumed in Table 3). Unlike the giant flares, the August 29 event did not show
an intense spike of hard photons with energies $\gtrsim 100$ keV. These pulses of $\gamma$-rays give evidence
for a vigorous relativistic outflow during the first 0.2 to 0.4 seconds of each giant outburst. The radio
afterglow detected following the August 27 event provides independent evidence for the ejection of particles
(\markcite{FKB99}Frail, Kulkarni \& Bloom 1999). However, the energy released on August 29 seems to have
been insufficient to drive such an outflow. On the basis of its BATSE light curve and spectrum, the bright
component of the August 29 event would be classified as a rare, unusually long SGR burst (non-flare), albeit 
with a fluence near the upper limit detected in ordinary bursts. The energies of these events span more than 
3.5 orders of magnitude, and follow a power-law distribution (\markcite{Gog99}{G\"o\u{g}\"u\c{s}} et al.~1999;
{G\"o\u{g}\"u\c{s}} et al.~2000).

On the basis of physical similarities, the main component of the August 29 event is listed in the same row
(``Bright X-ray Emission'') of Table 3 as the bright oscillating X-ray tails of the two giant flares.
These components all have hyper-Eddington luminosities ($L \sim 10^4 L_{Edd}$), similar quasi-thermal spectra
(OTTB temperatures $\sim 20-30$ keV), and little observed spectral evolution even as the flux declines
significantly. In each case, the emission terminates abruptly: after 3.5 s on August 29 (Fig.~1b and 1c) and after
about 370 s on August 27 (Feroci et al.~2000). (The final portions of the 1979 March 5 event were not clearly
observed.) In the magnetar model, these bright X-ray emissions come from a reservoir of hot, thermal
pair-photon plasma that is trapped on closed field lines in the magnetosphere (Thompson \& Duncan 1995). The
abrupt termination gives evidence for complete self-annihilation and evaporation of this plasma via
surface X-ray emission in a finite time (Feroci et al.~2000; Thompson et al.~2000b).

The main {\em difference} between the August 29 event and the giant flares is that its bright component was
too short to show a clear rotational modulation. The 3.5 s duration of this component was less than the 5.16
s spin period of SGR~1900+14, indicating that the trapped fireball evaporated before the star completed one
rotation.

The extended afterglow tail of the August 29 burst {\em does} show a clear rotational modulation (Fig.~4a and 5a), 
but its luminosity is orders of magnitude lower than those which were measured in the bright tails of the giant 
flares (or in the bright component of the August 29 event). Another interesting difference is the simple pulse profile of
the afterglow tail pulsations in the August 29 event versus a complex, evolving pulse shape in the August 27
and March 5 flares. The pulse profile of the August 29 tail did not change with time (over $\sim 1000$ s) and
we found no evidence for a 1-s sub-pulsation similar to that reported by Feroci et al.~in the tail of the
August 27 flare (\markcite{Fer99}Feroci et al.~1999). Furthermore, the spectrum of the August 29 afterglow
tail is quite distinct from those of the giant flare tails. The August 29 tail shows a significant 
softening over time, and its light curve declines gradually without any abrupt termination. For these reasons we 
conjecture that the August 29 afterglow tail {\em is a new component of SGR emission that has not been previously
observed or identified}, and is generated via some novel physical mechanism (or in a different location) than
the bright components of SGR outbursts. It should be noted that the published observations of the two giant
flares do not significantly constrain the presence of an extended afterglow tail, after the bright magnetospheric
emission has terminated. 

Based on observational ground, we {\it can} rule out extended afterglow tails following other ``ordinary" 
(non-giant flare) SGR bursts observed by RXTE. For example, among over 800 events from SGR~1900+14 and 
SGR~1806--20, no burst other than August 29 shows a modulated afterglow tail that lasts for longer than one 
rotational period. It was however predicted by the magnetar model (\markcite{TD95}Thompson \& Duncan 1995) 
that most SGR bursts show faint transient afterglows on a timescale comparable to the burst peak duration. 
Such short afterglows are mainly due to passive cooling from the neutron star surface, a mechanism that cannot 
explain the extended afterglow tail of August 29. This feature has been observed in many SGR bursts. For example 
Strohmayer and Ibrahim 1997 show a typical burst from SGR~1806--20 with peak and faint tail durations of about 
0.2 s and 0.3 s respectively. Bursts from SGR~1900+14 also show this behavior both before and after the August 29 
event.

It is difficult to make detailed comparisons between the precursors of the three events of Table 3, because
the events were studied with very different instruments. No precursor was detected in the Venera data before
the March 5 flare (Mazets et al.~1999a), but the detection threshold was relatively high because of
instrumental limitations and the remoteness of SGR~0526--66 in the Large Magellanic Cloud. A 
precursor was detected about $0.45$ s before the sharp onset of the August 27 event's hard spike.
It was a simple pulse about 0.05 s in width, evident in only the lowest-energy channel of the Konus WIND
experiment ($15-50$ keV; Mazets et al.~1999a) and by the GRB monitor aboard Ulysses ($25-150$ keV; Hurley et
al.~1999a). This precursor clearly had a softer spectrum than the $\gamma$-ray spike which followed, but it
could have been spectrally similar to the bright X-ray emissions in all three events of Table 3 and to some
parts of the August 29 precursor. Note that Konus also revealed a second precursor on August 27: a faint
component of smoothly-intensifying X-rays during the last $\sim 0.08$ s before the sudden onset of the hard spike,
again detected in only the lowest-$E$ channel (Mazets et al.~1999a).


\section{Discussion}

The different spectral and temporal signatures of the precursor, main peak and afterglow tail of the August 
29 event suggest that these three components are produced by different emission mechanisms.
Our spectral results indicate that the spectrum is uniform and statistically unvarying
during during the 3.5 s main peak of the August 29 event. Such spectral uniformity of SGR bursts was first 
noted by \markcite{Maz82}Mazets et al.~(1982) in their analysis of the March 5 flare from SGR~0526--66.
It was also found by \markcite{K87}Kouveliotou et al.~(1987) and \markcite{F94}Fenimore et al.~(1994) 
in their analyses of the spectra of bursts from SGR~1806--20. In addition, SGR bursts of widely differing 
fluences emitted by the same source have been found to have similar spectra (\markcite{F94}Fenimore et al.~1994). 

However, recent evidence from RXTE observations of SGR~1806--20 indicates that modest spectral variations can occur 
during SGR bursts (see \markcite{SI97}Strohmayer \& Ibrahim 1997). Our results on the precursor and 
the afterglow tail emissions point out that strong spectral evolution can also occur in SGR outbursts. The evolution in 
both the precursor and tail is hard-to-soft where the temperature decreases by more than 50\% in both regions. The 
decline in $kT$ during the precursor is accompanied by a slower decrease in $n_H$. However, during the tail $n_H$ does 
not show the same trend.
Modest spectral evolution has recently been observed in the August 27 giant flare from SGR~1900+14. 
For example, spectral modulations with pulse phase as well as overall modest spectral softening have been
reported during this event by several researchers (see \markcite{H99a}Hurley et al.~1999a; 
\markcite{Fer99}Feroci et al.~1999; \markcite{Maz99a}Mazets et al.~1999a).

In the following, we discuss the physical mechanisms that could produce the components of the August 29 event 
(i.e. precursor, peak, and tail), and account for their spectral properties. In section 5.1 we describe how giant 
flares and regular SGR bursts are produced in the magnetar model. In section 5.2 we discuss in detail the afterglow 
tail of August 29 event and elaborate on the possible mechanisms that could power its extended emission. The precursor 
is discussed in section 5.3. We conclude with a summary and final remarks in sections 5.4 and 5.5.


\subsection{Giant Flares and Regular SGR Burst Emission}

According to the magnetar model detailed in 
\markcite{TD95}Thompson \& Duncan (1995), the giant flares of March 5 and August 27 
involve a readjustment of the stellar magnetic field on large scales of up to several km. 
A flare occurs when the field reaches a point of instability, gated by the rigid neutron star crust, and
relaxes to a lower energy state. The extreme energetic output of $\sim 10^{45}$ ergs (\markcite{DT92}Duncan \&
Thompson 1992) and the very fast rise time of the March 5 flare (\markcite{PAC}Paczy\'nski 1992) point to such a 
source of `clean' energy. Indeed, a magnetic field stronger than $\sim 10\ B_{\rm QED} = 4.4\times 10^{14}$ G 
contains enough energy to power $\sim 100$ giant flares over the lifetime of an SGR source, and appears
needed to explain the extreme peak luminosity of $\sim 10^6-10^7$ Eddington (\markcite{TH2000}Thompson 2000). The
elastic energy of the deformed crust is, in itself, probably insufficient to power a giant flare; but
the energy available in the pinned magnetic field is much larger. For this reason,
the outburst is conjectured to be a hybrid of an earthquake and a `solar flare', involving 
a large propagating fracture in the crust of the neutron star that induces rapid reconnection and
large-amplitude wave motions in its magnetosphere. The short SGR bursts do not necessarily involve such a large-scale
fracture: a localized yield of the crustal lattice with sufficiently large fault slippage driven by magnetic
stresses could account for their energetics.

The initial $\gamma$-ray spikes of the giant flares have been identified with the rapid deposition of
energy in a hot fireball, which blows open some of the closed magnetic field lines that are anchored in the
neutron star (\markcite{TD95}Thompson \& Duncan 1995). The $\sim 0.2-0.5$ s duration of the spike is
comparable to the time for a $\sim 10^{15}$ G magnetic field to rearrange material in the core and deep crust
of the neutron star. The hard spectrum combined with rapid time-variability (e.g. Feroci et al.~1999;
\markcite{Maz99b}Mazets et al.~1999b) directly points to bulk relativistic expansion -- similar to but on a
smaller scale than the classical gamma ray bursts (GRBs). The observation of a radio afterglow from the
August 27 flare supports this view (see \markcite{FKB99}Frail, Kulkarni \& Bloom 1999). In this model,
a portion of the dissipated energy remains trapped in the magnetosphere, in the form of
an optically thick, electron-positron plasma that is essentially baryon free. This `trapped fireball'
cools by X-ray emission from its surface for hundreds of seconds, and is identified with the extended
pulsating tail of the giant flares. A recent analysis of the August 27 light curve provides direct evidence 
for this cooling mechanism, which involves a gradual contraction of the fireball photosphere to a sharp 
termination (\markcite{Fer2000}Feroci et al.~2000). 


\subsection{The August 29 Afterglow Tail}

Searches for afterglow from the heated surface of the neutron star provide a direct test of the presence
of a trapped fireball (\markcite{TD95}Thompson \& Duncan 1995). A fraction $\sim 10^{-2}-10^{-3}$ of the 
fireball energy is conducted into the relatively cold outer crust of the neutron star over the observed 
duration of the hard X-ray outburst, if the surface magnetic field is stronger than $\sim 10^{14}$ G. 
After the fireball dissipates, most of this energy will be conducted back out to the surface.
The luminosity of this afterglow radiation is correspondingly reduced with respect to 
the main burst. The internal fireball temperature is estimated to be $\simeq 1$ MeV in the giant flares, 
given a confinement volume of $\sim (10~{\rm km})^3$. 

In the short SGR bursts, the confining volume could be smaller, and we wish to constrain it in the case of 
the August 29 event. One infers $T\simeq 100$ keV for the August 29 burst if it was powered by a trapped 
fireball of similar dimensions to the August 27 giant flare. (Given the overall reduction of 
$\varepsilon_X = 0.004$ in fluence and the thermodynamic properties of a pair-photon plasma in a very 
intense magnetic field; \markcite{TD95}Thompson \& Duncan 1995.) However, the shorter duration and simpler 
light curve of the August 29 burst (without the conspicuous four-pronged profile of the giant flare) suggest 
a different geometry for this outburst. As we now describe, its afterglow, pulsating tail provides direct evidence 
for a small emitting area.

If we identify a surface hotspot with the pulsating tail, then the best-fit blackbody temperature of Table 2 
points to a radiative area of $2.2\,(1-2\,GM_{NS}/R_{NS}c^2)\,(D/7~{\rm kpc})^2$ km$^2$, less than one percent of the 
surface area of a neutron star. The factor $(1-2\,GM_{NS}/R_{NS}c^2) \simeq 0.6$ accounts for the gravitational 
redshifting of the measured temperature and luminosity. A bundle of closed magnetic field lines with this 
cross-sectional area and a length comparable to the stellar radius has a volume of $\sim 10$ km$^3$, and the 
fireball temperature is inferred to be $\sim 1$ MeV, similar to the giant flare. The duration of the main 
August 29 burst relative to the giant flare then depends on the geometry of the fireball: the duration is 
smaller by $\varepsilon_X = 0.004$ in planar geometry, and by $\varepsilon_X^{1/2}$ in cylindrical geometry. 
Since the relative durations\footnote{Of the components of the two bursts with weak measured spectral evolution.} 
are $3.5/400 \simeq 0.009$, we infer that the geometry of the August 29 fireball is closest to planar, e.g., 
that the plasma-loaded magnetic field lines straddle an extended fault. 

The light curve of the August 29 burst is also consistent with a planar geometry. A trapped fireball
cools as its outer surface contracts. The surface X-ray flux from a homogeneous fireball will have a 
flat-topped profile (and a very sharp termination) in planar geometry; whereas in cylindrical geometry the 
profile will be triangular and the termination more gradual.\footnote{A number of short SGR bursts are 
observed to have triangular profiles: see, e.g., \markcite{Maz99b}Mazets et al.~(1999b).} The pulsating 
phase of the August 27 giant flare, which involved a much larger disturbance of the magnetosphere, appears
 to be best fit by a fireball of approximately spherical geometry (\markcite{Fer2000}Feroci et al.~2000).


\subsubsection{Photospheric Expansion}

Can passive cooling of the heated surface of a neutron star account adequately for the total fluence and long 
duration of the August 29 tail emission? In the absence of photospheric expansion, most of the absorbed heat 
should be reradiated on the same $\sim 4$ s timescale over which the crust was exposed to a trapped fireball
(\markcite{TD95}Thompson \& Duncan 1995). 

Expansion is, however, inevitable when the fireball temperature is as high as $\sim 1$ MeV. During the main 
part of the burst (the "Bright X-ray Emissions" phase of Table 3) the trapped fireball {\it compresses}
that part of the stellar surface which lies beneath it. For a surface magnetic field $B > 10^{15}$ G, the 
relativistic Landau level excitation energy is $(2\,\hbar\,c\,e\,B)^{1/2} > 3\ kT$, so that only the lowest 
(one-dimensional) Landau state is populated. The fireball pressure at the star's surface is dominated by this 
relativistic, non-degenerate ($\mu_{e^\pm}=0$) pair gas:
$P_{e\pm} = {1\over12} eB(kT)^2/(\hbar\,c)^2$. 
In the cooler layer below, the electrons are compressed into a degenerate,
relativistic gas with a 1-D fermi energy $\mu_{e^-} = \pi kT/\sqrt{3}$.
This compressed layer is then heated, via radiative
diffusion, down to an electron column density
\begin{displaymath}
N_e \sigma_T \simeq 2 \times 10^9 (t_{burst}/4\,s)^{1/2} (B/10^{15} G) ( T/MeV)^{1/2}
\end{displaymath}
(cf.~\S 4.1 of Thompson \& Duncan 1995).

Before being heated by the August 29 fireball, this
surface layer had a hydrostatic pressure at its base of
$P_{\rm hyd} = N_e m_p\,g.$ Comparing this with the fireball pressure,
one finds
\begin{displaymath}
{P_{\rm hyd}\over P_{e^\pm}} = 0.005\,\left({N_e\sigma_T\over 10^9}\right)\,
\left({T\over {\rm MeV}}\right)^{-2}\,\left({B\over 10^{15}~G}\right)^{-1}\
\end{displaymath}
at a surface gravity of $g = 2\times 10^{14}$ cm s$^{-2}$. As a result, substantial surface layer expansion and 
wind emission will occur immediately after a trapped pair-photon plasma dissipates above any area of the crust. 
This wind supplies an ion-electron plasma which dominates the scattering opacity of the cool ($T \sim 10-20$ keV) 
fireball photosphere (\markcite{TD95}Thompson \& Duncan 1995). The vertical scale height of the heated layer 
expands by a factor 
$\simeq 10^2\,(B/10^{15}~{\rm G})\,(T/{\rm MeV})^2\,(N_e\sigma_T/10^9)^{-1}$.
The net effect of this expansion is to increase the radiative cooling time of this layer by a factor $\sim 30$ 
with respect to the initial heating time of $\sim 4$ s.

The net energy radiated by the August 29 afterglow tail implies more stringent
constraints on passive surface cooling. The energy absorbed from the fireball
over a timescale 
$t_{\rm burst}$ is 
$\simeq N_e T = 4 \times 10^{27}(t_{\rm burst}/4~{\rm s})^{1/2}\,(B/10^{15}~{\rm G})$ $(T/{\rm MeV})^{3/2}$ 
erg cm$^{-2}$ per unit area;
whereas the output of the afterglow tail is inferred to be somewhat larger, 
$4.4\times 10^{38}~{\rm erg}/1.3~({\rm km})^2 \simeq 3\times 10^{28}$ erg cm$^{-2}$. 
This difference could be explained if the surface magnetic field is stronger than $10^{15}$ G, or if
the area of the hotspot has been underestimated.


\subsubsection{Thermonuclear Burning}

Given the high temperatures to which the surface of the neutron star appears to be exposed, it is also worth 
considering thermonuclear burning as a supplemental source of energy. Indeed, the energy released per nucleon 
burning hydrogen to helium is somewhat larger, $\sim 7$ MeV, than the energy absorbed from the fireball.
Helium is photodissociated above a temperature of $\simeq 1.0$ MeV in a mildly degenerate surface layer.
This critical temperature for photodissociation decreases to $\simeq 0.3$ MeV at a density of 1 g cm$^{-3}$. 
The long duration of fireball emission in the August 27 giant flare suggests that photodissociation occurred 
most effectively in that previous event. 

In order to supply hydrogen for later burning, there are two additional requirements: first, neutrons
must be effectively converted to protons through positron capture, $n + e^+ \rightarrow p + \bar\nu_e$; and,
second, the proton-neutron-pair plasma must cool off rapidly, before the proton excess is reduced through
electron captures and the nucleons are bound up in alpha particles. At a temperature of 1 MeV, the 
timescale for positron capture is $n_n|dn_n/dt|^{-1} \simeq 2\,(B/10^{15}\,{\rm G})^{-1}$ s.
(A strong surface magnetic field $B > 10^{15}$ increases the phase space of the positrons at a fixed 
temperature.) This capture time is much shorter than the observed duration of the giant flares, and is 
comparable to the width of the main X-ray pulse in the August 29 flare. The equilibrium proportions of 
neutrons and protons depend on the degree of degeneracy; they are $n_n/n_p\simeq 0.1$ at $T = 1$ MeV,
in a pair-dominated plasma with $n_p \ll n_{e^+}$ and $n_{^4He} \simeq 0$. (Because the background 
neutrino flux is negligible, these abundances depart from nuclear statistical equilibrium.) In the trapped 
fireball model, decompression of the heated surface layer occurs over a tiny fraction of the duration of 
the SGR outburst, as the thin radiative surface layer of the fireball contracts toward its center. 
By contrast, the time for electron captures to change the proton density is 
$\sim 20\,(B/10^{15}\,{\rm G})^{-1}$ s at $T\sim 1$ MeV, and becomes much longer as the temperature drops 
during decompression. This guarantees that a significant fraction of the hydrogen created by photodissocation
will be retained for subsequent burning.

The burning history of the heated layer, and the mechanism by which burning could be triggered at a subsequent 
outburst, will be explored elsewhere. Nonetheless, it should be emphasized that direct empirical evidence
for extended H-burning flares is present in a source of a very different nature, the recurrent transient 
Aql X-1. The first Type I X-ray burst detected during an outburst in March/April 1979 had a very long tail 
lasting some 2500 s (\markcite{Cze87}Czerny, Czerny, \& Grindlay 1987). It has been noted for independent 
reasons that a light hydrogen-helium atmosphere will increase the surface X-ray flux from a warm magnetar
(\markcite{Heyl98}Heyl \& Hernquist 1998, and references therein).

\subsubsection{Persistent Particle Flows and Non-thermal Spectra}

Are the hard spectrum and large-amplitude pulsations of the afterglow tail consistent with this 
interpretation? Can the X-ray photons detected above $\sim40$ keV survive splitting in the
intense magnetic field inferred for SGR~1900+14? Note, first, that the large black body temperature (Table 2) 
implies a radiative flux 4 to 20 times larger than the classical Eddington flux. The radiative flux (and hence 
the `Eddington' flux) can be increased by a suppression of the magnetic scattering opacity in the strong 
magnetic field (\markcite{PAC}Paczy\'nski 1992), although this effect occurs unambiguously only for radiative 
diffusion {\it across} a confining magnetic field  (\markcite{TD95}Thompson \& Duncan 1995). The radiative force 
on matter near the neutron star surface is increased through several effects, including conversion of the two 
polarization modes by Compton scattering below the electron cyclotron resonance (\markcite{Mill95}Miller 1995; 
\markcite{TD95}Thompson \& Duncan 1995) and near the proton cyclotron resonance (\markcite{TH2000}Thompson 2000). 
Indeed, the inferred radiative flux of the afterglow tail is sufficient to lift protons off the stellar surface 
through scattering at the proton cyclotron line (energy $\hbar\,e\,B/m_p\,c = 6.3\,(B/10^{15}~{\rm G})$ keV, and 
integrated scattering cross section $(\pi/4\alpha_{\rm em})(B/B_{\rm QED})^{-1}\sigma_T$ in a dipole magnetic field). 
The critical isotropic X-ray luminosity is $E L_E \geq 1.2\times 10^{37}\,(B/10^{15}~{\rm G})$ erg s$^{-1}$ at the 
line. As the protons (and neutralizing electrons) accelerate from the surface, the range of frequencies
that interact resonantly with the line becomes Doppler-broadened, and a significant fraction of the radiative flux 
could be converted to bulk kinetic energy. Material excavated from depth will carry heavier elements processed by 
nuclear burning.

The presence of such a particle flow opens up the possibility of creating a non-thermal spectrum
through Comptonization of the {\it ordinary} polarization mode. This mode is the dominant coolant for hot
electrons because it has a large scattering cross section (close to Thomson) and because it 
{\it does not} split even in magnetic fields much stronger than $B_{\rm QED}$. (Only the orthogonal 
extraordinary mode can split.) An added bonus is that the emergent O-mode radiation has a tendency
to be beamed along the magnetic field lines (\markcite{Bas75}Basko \& Sunyaev 1975), because the 
net scattering cross-section of this mode varies as $\sin^2\theta$ with the angle $\theta$ between
the incident photon and a very strong
magnetic field (e.g. \markcite{Her79} Herold 1979).

Detection of faint afterglow radiation from the heated surface of a strong-B neutron star, containing
both blackbody and power-law components in its spectrum, has potentially interesting implications for the 
persistent emission of the Soft Gamma Repeaters and especially the Anomalous X-ray Pulsars. The non-thermal 
emission of SGR~1900+14 brightened by a factor $\sim 2.5$ and simplified into a single pulse following the 
August 27 giant flare (\markcite{Mur99}Murakami et al.~1999; Woods et al.~1999). This effect has been ascribed, 
in the magnetar model, to a persistent current driven by a twisting up of the external magnetic field lines during
the August 27 giant flare (\markcite{TH2000a}Thompson et al.~2000a). 


\subsubsection{Alternative Mechanism: Ejection and Delayed Fallback}

An alternative model for the afterglow radiation of the August 29 burst should be mentioned. During an SGR 
outburst, a modest amount of material can be ejected beyond the corotation radius 
$R_{\rm co} = (GM_{\rm NS})^{1/3}(P_{\rm rot}/2\pi)^{2/3}$, where it is centrifugally supported, and 
(temporarily) confined there by magnetic tension: 
$\Delta M \sim B_{\rm dipole}^2 R_{\rm NS}^6\Omega^{4/3}/4\pi(GM_{\rm NS})^{5/3}
= 2\times 10^{20}\,(B_{\rm dipole}/4\times 10^{14}~{\rm G})^2\,
(P_{\rm rot}/8~{\rm s})^{-4/3}$ g (\markcite{W2000}Woods et al.~2000).
The corresponding column density (assuming this material to be mainly hydrogen) is 
$N_H \sim 3\times 10^{25}\,(B_{\rm dipole}/4\times 10^{14}~{\rm G})^2\,(P_{\rm rot}/8~{\rm s})^{-8/3}$ cm$^{-2}$.
After an SGR outburst, this suspended material will cool off and settle into a thin, rotationally-supported disk. 
As this disk thins out, the centrifugal force density rises with respect to $B_{\rm dipole}^2/4\pi$ at the 
corotation radius, and the disk begins to spin outward adiabatically. 

During a subsequent SGR outburst this material will be re-heated, and some may spill back across the corotation 
radius, where it is no longer centrifugally supported against gravity and can collapse back onto the neutron star. 
The covering fraction $\Delta\Omega_{\rm spot}/4\pi$ of the resulting surface hotspots depends on the geometry of 
the magnetic field. However, in a pure dipole geometry it is too small: 
$\Delta\Omega_{\rm spot}/4\pi = ({1\over 4}-{1\over 2}) (R_{NS}/R_{co})
\sim 4-7\times 10^{-4}\,(P_{\rm rot}/8~{\rm s})^{-2/3}$.
We are not aware of a simple argument leading to accretional luminosity that is $\sim 10^{-2}-10^{-3}$ of the SGR 
burst luminosity. Nonetheless, the energetics are acceptable for the August 29 afterglow tail: a net accretional energy 
up to $\sim 10^{40}$ erg could be released. It should also be emphasized that this process can make only a tiny 
contribution to the extended persistent emission of the SGR sources.


\subsection{The Precursor Emission}

The precursors detected before the August 27 and 29 outbursts offer an interesting test of the idea that SGR
bursts arise from a trapped, pair-loaded plasma. The weak precursor of August 27 did not yield a meaningful 
spectrum and no spectral properties were reported; however, it was estimated that it had a much softer
spectrum than the rest of the event (\markcite{Maz99a}Mazets et al.~1999a). The precursor of August 29
event, as observed by RXTE, has several interesting features. It is relatively long ($\sim$ 1 s), has complex
structure with multiple peaks, and showed significant spectral evolution. The measured fluence of short SGR 
outbursts covers a very wide range of up to four decades (\markcite{Gog2000}{G\"o\u{g}\"u\c{s}} et al.~2000),
and indeed the precursor lies within this established range. The individual features have durations $\sim 0.1$ s
 not atypical of SGR bursts; the light curve is unusual in that the several X-ray pulses are connected
up into a continuous period of emission.

The spectrum at the beginning of the precursor is harder than any other point during the first 
$\sim4.5$ s of the event, including the burst peak. The hardness ratio (Fig.~3b) shows a systematic softening of 
the spectrum in the first 0.5 s. The precursor also showed a very fast rise time of 9.8 ms at its onset, which is 
comparable to the $\sim 4$ ms rise time of the intense $\gamma$-ray spike of August 27 flare. We also
notice that although there is no clear temporal distinction between the end of the precursor and the
beginning of the main peak, they have different spectral properties. 

In the trapped fireball model for SGR outbursts, the thermodynamic properties of the emitting plasma in a faint 
outburst depend crucially on the geometry. The rapid injection of a small amount of energy in a correspondingly
small volume (involving e.g. a localized adjustment of the magnetic field lines over a small patch of the neutron 
star surface), will trigger the formation of a trapped thermal fireball. However, if the energy is injected more 
gradually over a larger volume, then it is possible to establish a continuous balance between electrostatic 
heating of the suspended pairs, and their diffusive radiative cooling. The critical rate of energy injection 
(through reconnection and dissipation of charge-starved Alfv\'en waves) is 
$L_{\rm crit} \sim 10^{42}(L/10~{\rm km})$ erg s$^{-1}$,
where $L$ is the characteristic dimension of the heated magnetospheric plasma (idealized as being spherical in this 
analysis; \markcite{TDFH2000b}Thompson et al.~2000b). Below this injection luminosity, the heated O-mode photons 
can maintain an approximately Wien distribution at the same temperature as the pairs (approximately 20 keV in the 
plasma interior), with a stable balance between diffusive loss and creation by splitting. Above this injection 
luminosity, the photons are approximately black body, and the plasma is unstable to an upward perturbation to the 
pair density and temperature. This causes a runaway to a very dense, hot fireball which cools on a much longer 
timescale, via a diffusive surface cooling wave (\markcite{TD95}Thompson \& Duncan 1995). The initial luminosity of 
the giant flares is measured to exceed $L_{\rm crit}$, which together with the rapid termination of the bursting
flux (\markcite{Fer2000}Feroci et al.~2000) directly points to the formation of a dense, hot fireball. However, the 
intermediate $\sim 40$ s of the August 27 flare, before the appearance of large-amplitude pulsations, had a somewhat 
harder spectrum and has been identified with continuing seismic input leading to the formation of an extended pair 
corona (\markcite{Fer2000}Feroci et al.~2000).

The rapid rises of many short SGR bursts are also consistent with a rapid injection of energy into the
magnetosphere, on a timescale much shorter than the observed X-ray outburst. The main 3.5 s component of the
August 29 burst appears to fit this description. However, the relatively high temperature measured in the
first peak of the August 29 precursor, its relatively low peak luminosity ($\sim 10^{-2}$ of the main burst),
and the long duration of the precursor, are all suggestive of a more gradual energy input, below the
luminosity $L_{\rm crit}$. The high temperature is consistent with bounds from photon splitting,
if a significant component of the radiative flux is carried by the Compton-heated O-mode (which does not
split) and/or if the observed high-energy photons are produced outside the spitting photosphere at $B \sim
B_{\rm QED}$. Some previously analyzed SGR outbursts with hard spectra (Strohmayer \& Ibrahim 1997) may also
involve such a radiative mechanism. 


\subsection{Summary}

As discussed above, the afterglow tail emission of the August 29 event, and its spectral evolution, could be explained by a 
hot spot that covers $\sim 1$ percent of the neutron star surface. In the trapped fireball model for SGR bursts (including the 
main 3.5 s component of August 29), such a hotspot is predicted to form when a small patch of the neutron star crust is exposed 
to high temperatures ($T\sim 1$ MeV). 

Although vertical expansion of this heated surface layer will prolong the cooling X-ray flux, the measured fluence 
of the afterglow tail may point to an additional energy source: we suggest burning in a surface layer of hydrogen and helium 
that results from photodissociation by SGR burst fireballs. Thus the oscillatory tails of the August 27 and 29 bursts are ascribed 
to independent mechanisms: in the giant flare, the envelope of the pulsations is consistent with a cooling, trapped fireball 
(\markcite{Fer2000}Feroci et al.~2000). It should be emphasized that published observations of the August 27 giant flare do not 
presently constrain the presence of a afterglow tail, formed by surface heating, following the termination of the magnetospheric 
emission.

The precursor and main peak of the August 29 burst appear both to involve emission from a trapped plasma, but with different 
emission properties, that on one hand produce a spectrally evolving precursor, and on the other hand a spectrally uniform 
burst peak. The rate of energy injection in the precursor may be below the critical value for the formation of a truly thermal 
fireball, thereby allowing a more direct balance between heating and cooling. 


\subsection{Conclusion}

The August 29 event is unique amongst SGR bursts. The unusually long $3.5$ s duration of its peak emission, 
the presence of an extended precursor, and the very extended periodic tail, distinguish it from ordinary 
SGR bursts. The spectral signatures seen in this event are also quite remarkable. Recently, a 6.4 keV emission line 
has also been discovered in the precursor of this event (\markcite{SI2000}Strohmayer \& Ibrahim 2000). This is the first 
ever emision line to be detected from an SGR.

While the shape of the light curve has some resemblance to the giant flares of March 5 and August 27, 
the August 29 event released much less energy and did not show the initial, hard $\gamma$-ray
spike seen in giant flares. The luminosity and spectrum of its main pulse were, in fact, much closer
to the luminosity and spectrum of the {\it pulsating tail} of the August 27 flare that
preceded it. Nonetheless, both bursts from SGR~1900+14 were initiated by a precursor,
which points to a basic similarity in the triggering mechanism. 

The occurrence of the August 29 event less than two days after the August 27 giant flare, its relatively large fluence and 
unusually long durations (by the standards of ordinary SGR bursts), and the appearance of a precursor in both outbursts, 
suggests that the August 29 event is an `aftershock' from the August 27 giant flare. 

\section{Acknowledgments}

A. I. is grateful to Jean Swank for carefully reading the manuscript and many useful advice, and to 
David Palmer, Samar Safi-Harb, and Craig Markwardt for many helpful comments. He also wishes to thank 
Kevin Hurley for providing the Ulysses data, which allowed looking at the event with three different 
instruments. P. M. W. and C. K. acknowledge support under the LTSA grant NAG 5-9350. C. T. acknowledges
support from NASA grant NAG5-3100 and the Alfred P. Sloan foundation. R. D. acknowledges support from NASA
(NAG5-8381) and the Texas Advanced Research Project (ARP-028).



\begin{deluxetable}{cccccc} 
\tabletypesize{\small} 
\tablecolumns{5} 
\tablewidth{0pt}
\tablecaption{Spectral results for the precursor, burst, and tail with the OTTB model}
\tablehead{ 
\colhead{Interval} & \colhead{$kT$ (keV)} &\colhead{$n_H\ (10^{22}$ cm$^{-2}$)} &\colhead{Flux ($10^{-8}$ ergs cm$^{-2}$ s$^{-1}$)} & 
\colhead{$\chi^2_{\nu}$}}
\startdata 
Precursor 1 & 28.72 $\pm$ 4.3 & 5.21 $\pm$ 0.6 & 6.34 & 1.15 \\
Precursor 2 & 18.56 $\pm$ 2.1 & 5.13 $\pm$ 0.5 & 4.55 & 1.00 \\ 
Precursor 3 & 14.82 $\pm$ 1.5 & 4.72 $\pm$ 0.5 & 6.14 & 1.06 \\ 
Precursor 4 & 11.79 $\pm$ 1.0 & 3.81 $\pm$ 0.4 & 7.72 & 0.60 \\
Precursor 5 & 19.54 $\pm$ 2.1 & 5.42 $\pm$ 0.6 & 2.57 & 0.95 \\ \tableline 
Burst rise  & 20.0 $\pm $ 2.3 & 3.3 $\pm$ 1.9 & 12.28 & 0.82 \\ 
Burst fall-off & 20.9 $\pm$ 2.1 & 5.48 $\pm$ 1.8 & 24.35 & 0.95 \\ \tableline 
Tail 1 & 73.2 $\pm$ 19.7 & 3.51 $\pm$ 1.0  & 0.287  & 1.08 \\ 
Tail 2 & 42.0 $\pm$ 8.2  & 8.25 $\pm$ 1.4 & 0.0911 & 0.94 \\ 
Tail 3 & 57.7 $\pm$ 17.3 & 7.52 $\pm$ 1.5 & 0.0574 & 1.56 \\ 
Tail 4 & 36.9 $\pm$ 8.5  & 10.5 $\pm$ 2.0  & 0.0376  & 1.56 \\ 
Tail 5 & 29.4 $\pm$ 6.8  & 8.81 $\pm$ 2.0  & 0.0297  & 1.52 \\ 
Tail 6 & 27.3 $\pm$ 5.9  & 9.0  $\pm$ 1.8  & 0.0277  & 1.15 \\ 
Tail 7 & 29.1 $\pm$ 6.0  & 7.1  $\pm$ 1.5  & 0.0218  & 1.09 \\ 
Tail 8 & 13.7 $\pm$ 1.8  & 13.1 $\pm$ 1.5  & 0.017   & 1.68\\  \tableline
\enddata 
\end{deluxetable}


\begin{deluxetable}{ccccccc} 
\tabletypesize{\small} 
\tablecolumns{7} 
\tablewidth{0pt}
\tablecaption{Spectral results for the tail with (PL+BB) model} 
\tablehead{ 
\colhead{Interval} & \colhead{$kT$ (keV)} & \colhead{$\alpha$ \tablenotemark{(a)}} & 
\colhead{$f_{tot}$ \tablenotemark{(b)}} & \colhead{$f_{BB}$ \tablenotemark{(c)}} & \colhead{$f_{PL}$ 
\tablenotemark{(d)}} & \colhead{$\chi^2_{\nu}$} }

\startdata 

1 & 4.24 $\pm$ 0.53 & 1.61 $\pm$ 0.23 & 6.02 $\times 10^{-9}$ & 2.4 $\times 10^{-9}$ & 3.7 $\times 
10^{-9}$ & 0.86 \\

2 & 4.04$\pm$ 0.48 & 1.98 $\pm$ 0.51 & 19.6 $\times 10^{-10}$ & 8.7 $\times 10^{-10}$ & 10.9 $\times
10^{-10}$ & 0.84 \\

3 & 4.04 $\pm$ 0.33 & 4.04 $\pm$ 1.1 & 12.9 $\times 10^{-10}$ & 9.7 $\times 10^{-10}$ & 3.2 $\times
10^{-10}$ & 1.22 \\

4 & 3.46 $\pm$ 0.26 & 1.65 $\pm$ 0.7 & 8.5 $\times 10^{-10}$ & 6.1 $\times 10^{-10}$ & 2.4 $\times
10^{-10}$ & 1.12 \\

5 & 3.41 $\pm$ 0.25 & 3.17 $\pm$ 1.1 & 6.1 $\times 10^{-10}$ & 5.0 $\times 10^{-10}$ & 1.12 $\times
10^{-10}$ & 1.06 \\

6 & 3.66 $\pm$ 0.25 & 2.1 $\pm$ 0.5 & 5.4 $\times 10^{-10}$ & 1.9 $\times 10^{-10}$ & 3.4 $\times
10^{-10}$ & 1.18 \\

7 & 3.37 $\pm$ 0.27 & 2.2 $\pm$ 0.8 & 4.4 $\times 10^{-10}$ & 2.6 $\times 10^{-10}$ & 1.8 $\times
10^{-10}$ & 0.94 \\

8 & 2.73 $\pm$ 0.22 & 4.2 $\pm$ 1.3 & 3.2 $\times 10^{-10}$ & 2.4 $\times 10^{-10}$ & 0.6 $\times
10^{-10}$ & 1.38 \\  \tableline

\enddata
\tablenotetext{(a)} {Power Law Photon index} 
\tablenotetext{(b)} {Total flux} \tablenotetext{(c)}
{Blackbody flux} \tablenotetext{(d)} {Power law flux, all in ergs cm$^{-2}$\ s$^{-1}$}
\end{deluxetable} 



\begin{deluxetable} {cp{0.5in}p{1.15in}p{1.4in}p{1.3in}p{1.3in}} 
\tabletypesize{\small}
\tablecolumns{5} 
\tablewidth{0pt} 
\tablenum{3}
\tablecaption{Comparision Between August 29 Event and August 27 and March 5 Giant Flares}
\tablehead{ 
\colhead{} & \colhead{} & \colhead{} & \colhead{\bf{August 29 Event\tablenotemark{(a,b)}}} & 
\colhead{\bf{August 27 Flare\tablenotemark{(c,d,e)}}} & 
\colhead{\bf{March 5 Flare\tablenotemark{(f,g,h)}}} }

\startdata 

	        & \multicolumn{2}{l}{Source}              & SGR~1900+14         & SGR~1900+14 & SGR~0526--66 \\
{\bf Global}    & \multicolumn{2}{l}{Distance (kpc)}      & 7              & 7      & 55  \\
{\bf Properties}& \multicolumn{2}{l}{SNR}                 & G42.8+0.6           & G42.8+0.6   & N49 in LMC  \\
                & \multicolumn{2}{l}{Total Energy (ergs)} & $1.1\times 10^{41}$ & $\gtrsim1.2\times 10^{44}$ & $\gtrsim5.2\times 10^{44}$ \\ \cline{1-6}
  	        & {\it Temporal} & {Rise time} & 9.8 ms                     & $\sim 18$ ms  & \\ 
 
		& {\it Behavior} & {Duration}  & 0.85 s                     & $\sim 0.05$ s &  \\ 

{\bf Precursor} &                & {Structure} & Complex, multipeaks    & simpler & \qquad None Observed\\ \cline{2-5}

	        & {\it Spectral} & {OTTB kT (keV)}      & $28.7\pm4.3\to11.8\pm1.0$  & $\sim20$     &   \\ 

	        & {\it Behavior} & {Spectral Evolution} & Significant (hard-to-soft) &  N/A         &   \\ \cline{1-6} 
  	        & 		 & {Rise time}     &                     & $<4$ ms  &  $<1$ ms      \\ 
 
		& {\it Temporal} & {Fall-off Time} &                 &  $\sim 35$ ms &  $\sim 40$ ms \\ 

{\bf Initial}   & {\it Behavior} & {Duration}      & 		 & $\sim 0.4$ s &  $\sim 0.2$ s \\ 

{\bf Hard }     &                & {Termination}   &                     & Gradual      & Gradual \\ \cline{2-3} \cline{5-6}        

{\bf $\gamma$-ray} & {\it Spectral} & {OTTB kT (keV)}      &  \qquad None Present & $kT\sim240$     & $kT\sim500$  \\ 

{\bf spike}        & {\it Behavior} & {Spectral Evolution}  &                &  N/A            &  N/A  \\ \cline{2-3} \cline{5-6}
		&       & \multicolumn{2}{l}{Flux (ergs cm$^{-2}$ s$^{-1}$)} & $\gtrsim3.1\times 10^{-2}$ & $1\times 10^{-3}$ \\ 
		
	        & {\it Energetics}& \multicolumn{2}{l}{Luminosity (ergs s$^{-1}$)}& $\gtrsim3.7\times 10^{44}$ & $3.6\times 10^{44}$ \\ 

	        & 	& \multicolumn{2}{l}{Fluence (ergs cm$^{-2}$)}   & $\gtrsim5.5\times 10^{-3}$ & $4.5\times 10^{-4}$ \\ \cline{1-6} 

             &                &  {Rise Time} & 20 ms  & N/A   & N/A \\ 

             &                & {Fall-off Time} & 17 ms  & N/A & N/A \\ 

             & {\it Temporal}     & {Duration}      & 3.5 s  & $\sim 370$ s  & $> 140$ s \\ 

	     & {\it Behavior}     & {Pulse Period}        &  N/A   & 5.16 s  & 8.1 s \\ 

{\bf Bright}  &                & {Pulse Profile} &  N/A   &  Complex & Complex \\ 

{\bf X-ray}    &              & {Profile Evolution?} & N/A & Yes, four 1-s sub-peaks & Yes, 2--4 sub-peaks\\ 

{\bf Emissions}&                & {Termination}   & Very Sharp & Sharp &  N/A \\ \cline{2-6}
                & {\it Spectral} &  OTTB kT (keV) & $\sim 20.6$  & $34.2\to28.9$ & $\sim30$ \\ 

                & {\it Behavior} & {Spectral Evolution} & Insignificant & Modest & Modest \\ \cline{2-6}
		& {\it Energetics} & {Fluence (ergs cm$^{-2}$)} & $1.9\times 10^{-5}$  & $4.2\times 10^{-3}$ & $1\times 10^{-3}$ \\ 

                &    & {Energy (ergs)} & $1.1\times 10^{41}$ & $5.2\times 10^{43}$ & $3.6\times 10^{44}$ \\ \cline{1-6}             
		
		& 		 & {Duration}  & $>1000$ s  \\ 
             
		& 		 & {Pulse Period}  & 5.16 s   \\ 

		& {\it Temporal} & {Pulse Profile} & Simple, one peak  \\ 

{\bf Afterglow} & {\it Behavior }& {Profile Evolution?}  & No \\ 

{\bf Tail}      & 		 & {Tail Bursts} & Yes, several & \qquad None Observed & \qquad None Observed \\ 

{\bf Emissions}  & 		 & {Termination} &  Gradual  \\ \cline{2-4}
		& {\it Spectral} & {OTTB kT (keV)} & $73.2\pm19.7\to13.7\pm1.8$ \\

		& {\it Behavior} & {Spectral Evolution} & Significant (hard-to-soft) \\ \cline{2-4}		
		& {\it Energetics} & {Fluence (ergs cm$^{-2}$)} & $7.5\times 10^{-8}$  \\ 
		
		&   & {Energy (ergs)}  & $4.4\times 10^{38}$ & \\ \tableline






\enddata

\tablenotetext{(a)}{ RXTE: This work} 
\tablenotetext{(b)}{ BATSE: This work} 
\tablenotetext{(c)}{ Ulysses: Hurley et al.~1999a}
\tablenotetext{(d)}{ Konus: Mazets et al.~1999a}
\tablenotetext{(e)}{ BeppoSAX: Feroci et al.~1999; Feroci et al.~2000}
\tablenotetext{(f)}{ ICE-PVO: Fenimore et al.~1996}
\tablenotetext{(g)}{ SIGNE II MS-Venra 12 \& Prognoz 7: Barat et al.~1979}
\tablenotetext{(h)}{ ISEE 3: Cline 1980}

\end{deluxetable} 

\clearpage

\clearpage


\begin{figure*}
\epsscale{1.0} 
\plotone{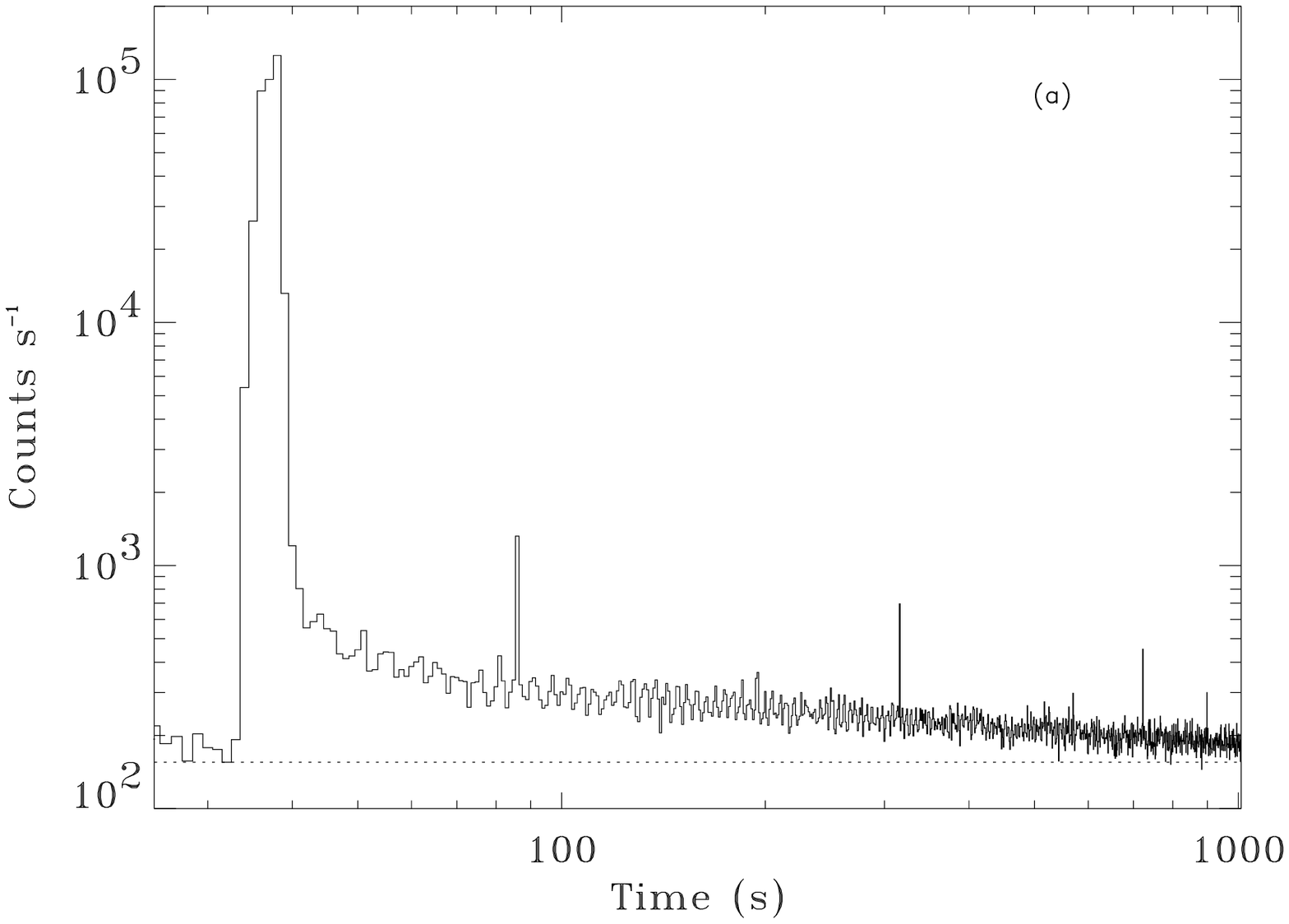} 
\end{figure*}

\begin{figure*}
\figurenum{1}
\epsscale{1.2} 
\plotone{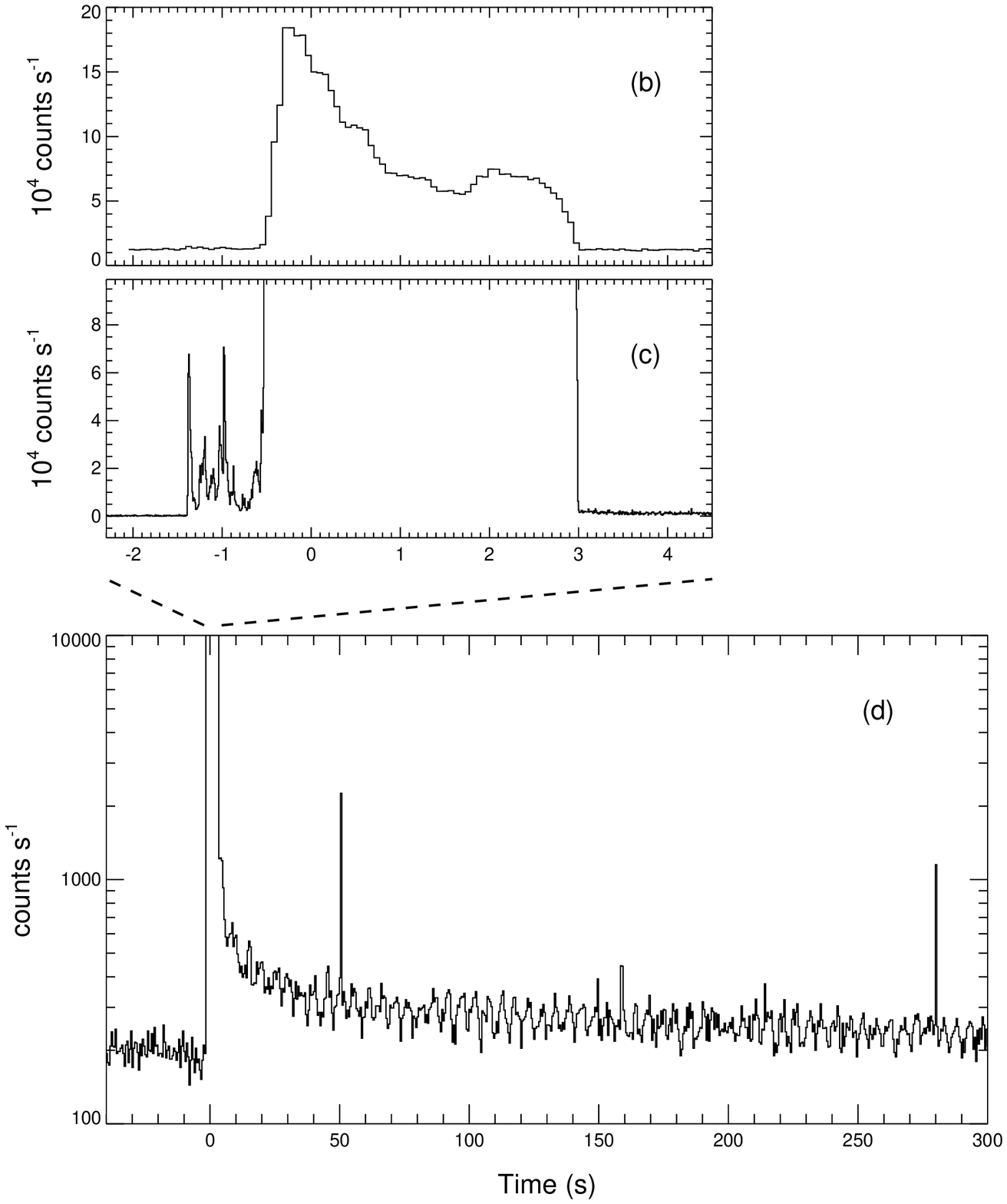} 
\caption{
Time history of the August 29, 1998 event from SGR~1900+14 as observed by RXTE/PCA (a, c, d) and BATSE (b). 
(a) In this log-log graph the long tail and the 5.16 s pulsations are clearly visible after 
the 3.5 s main burst peak. Several short recurrent bursts are seen during the tail. The precursor is not very 
visible here due to the low 1 s time resolution (see Fig.~1c and 3). The horizontal dotted line represents 
the background level. 
(b),(c) Simultaneous BATSE and RXTE light curves in 8 ms time resolution. The precursor is easily seen in the 
PCA profile, but barely discernible in the BATSE data. 
(d) The first 300 s of the event in a linear time scale.
$T_0=36992.481$ second of day (UT) in panels (b), (c).
} 
\end{figure*}

\clearpage

\begin{figure*}
\figurenum{2}
\epsscale{1.5} 
\plotone{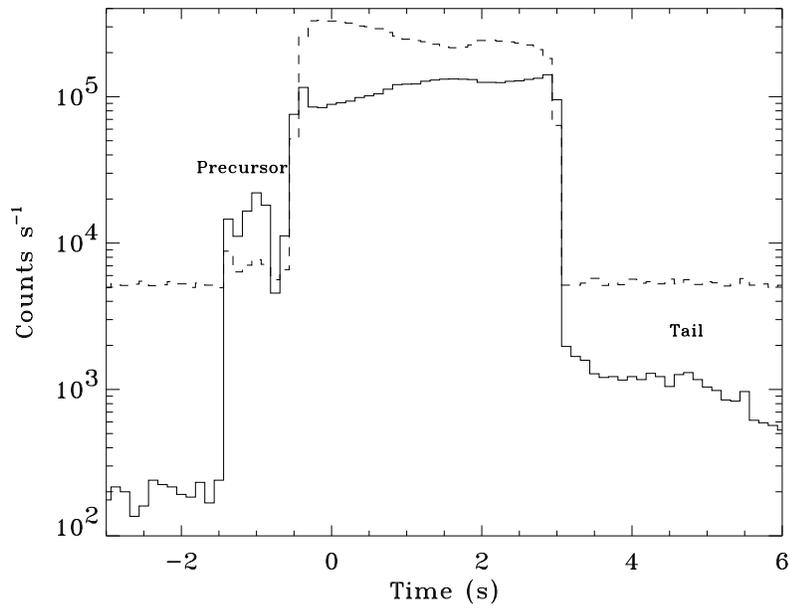} 
\caption{
Time history of the good X-ray (solid) and remaining counts (dashed) rates from the RXTE/PCA 
standard1 data mode during the August 29 event. During the main peak the remaining counts rate 
dominates and deadtime is severe. $T_0=36992.481$ second of day (UT).
} 
\end{figure*}

\clearpage

\begin{figure*}
\figurenum{3}
\epsscale{1.9} 
\plotone{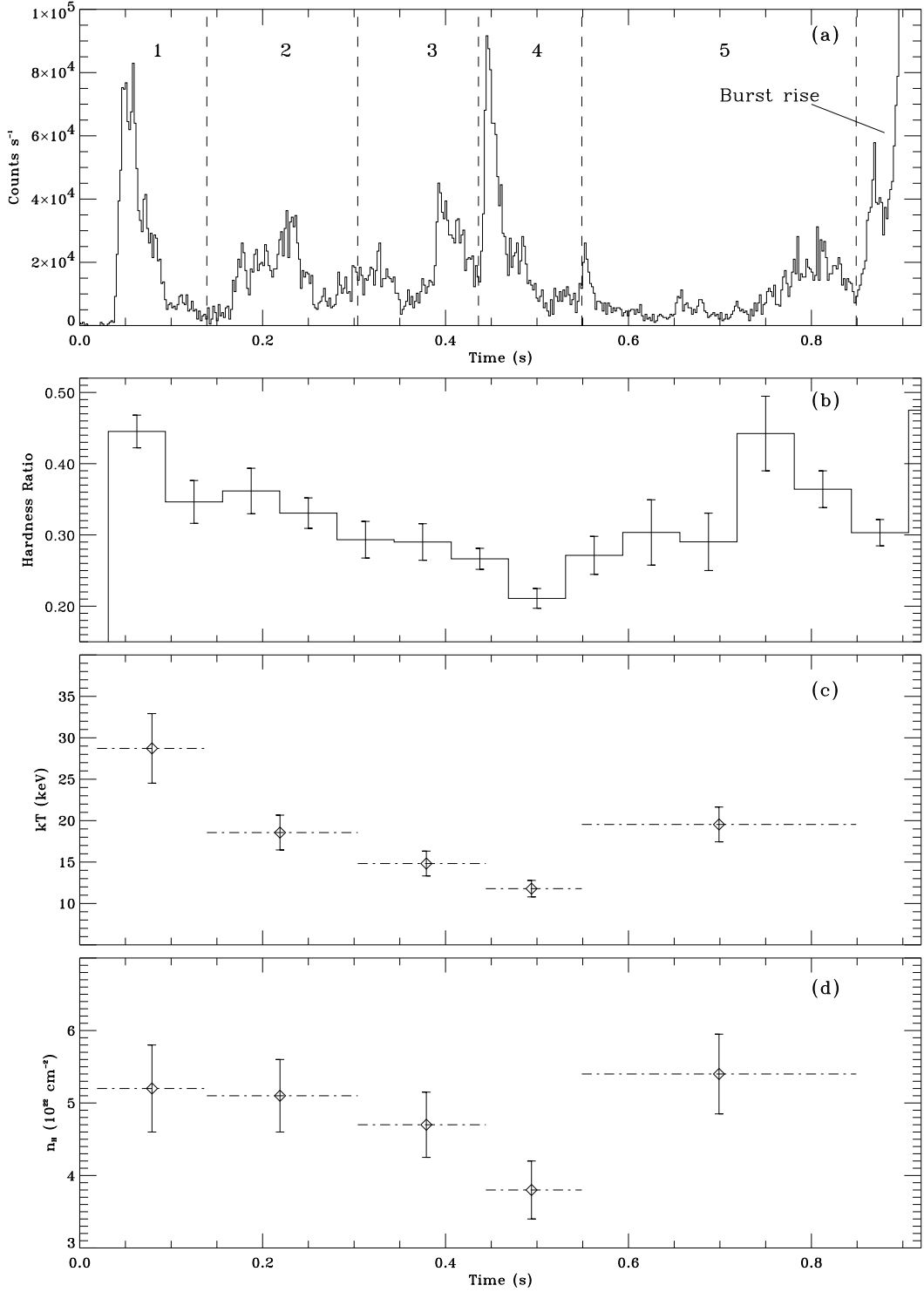}
\caption{
Temporal and spectral history of the precursor of the August 29 event with RXTE/PCA. 
(a) Precursor time history with the 5 intervals used for spectral analysis delineated by vertical dashed lines.
(b) The hardness ratio of the two energy windows ($11-50$ keV)/($2-10$ keV). 
(c) Time evolution of the OTTB temperature 
(d) Time evolution of the inferred column density of Hydrogen.
The horizontal dashed lines across the data points in panels (c) and (d) represent the intervals width.
} 
\end{figure*}

\clearpage

\begin{figure*}
\figurenum{4}
\epsscale{1.8} 
\plotone{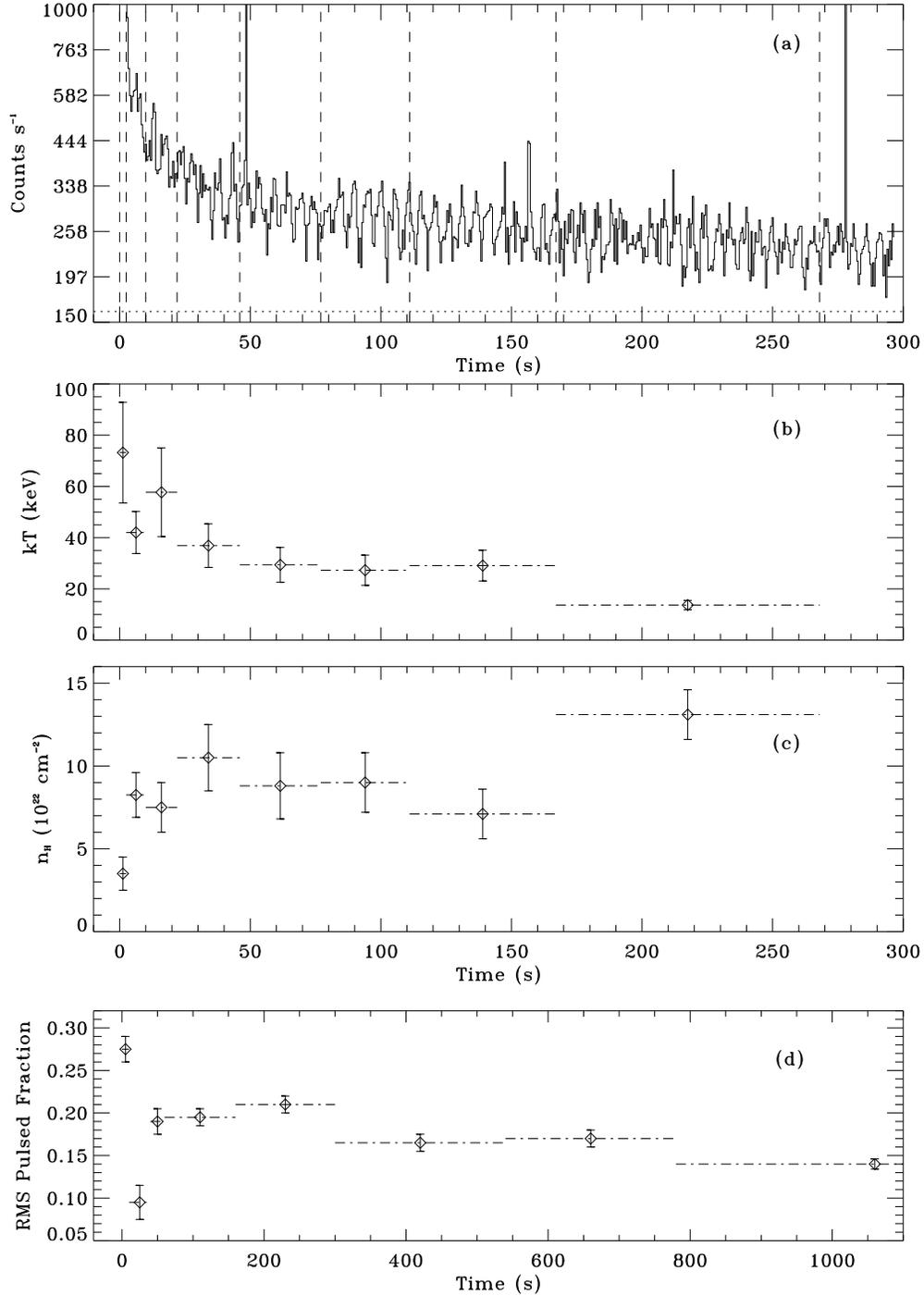} 
\caption{
(a) Time history of the first 300 s of the August 29 extended tail. The intervals used in the 
spectral analysis are denoted by the vertical dashed lines (tail bursts were removed before 
analyzing the data). The fine structure of the 5.16 s pulsations is clearly visible in this plot. The horizontal 
dotted line represents the background level. Note that the counts rate on the y-axis is plotted on a log scale. 
The next three panels show the time evolution of 
(b) the OTTB temperature,  
(c) the inferred column density of Hydrogen, and 
(d) the RMS pulsed fraction up to $t\sim1000$ s.
The horizontal dashed lines across the data points in panels (b), (c), and (d) represent the intervals width used 
in the analysis.
} 
\end{figure*}

\clearpage

\begin{figure*}
\figurenum{5}
\epsscale{1.8} 
\plotone{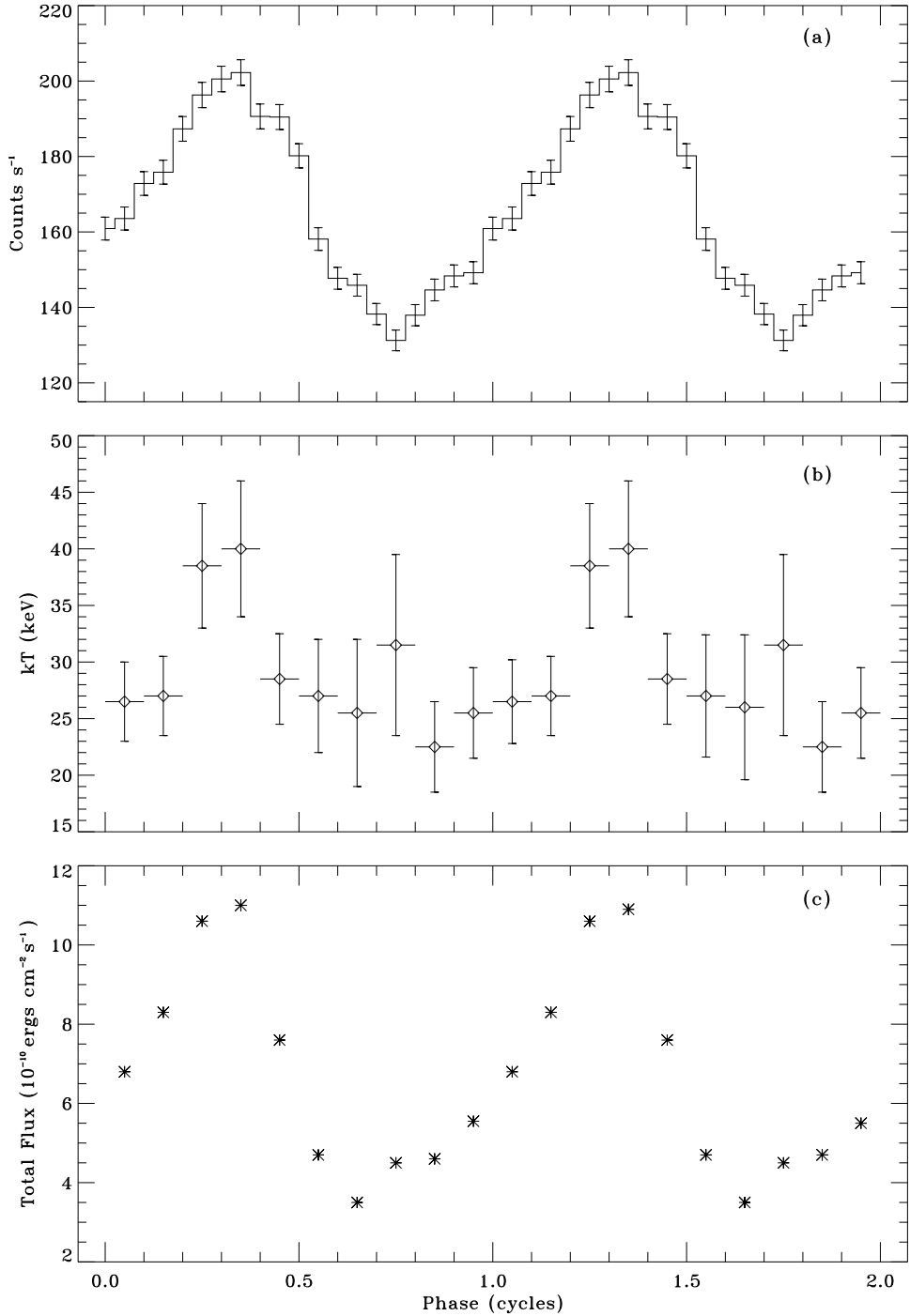} 
\caption{
(a) Phase folded pulse profile of the tail pulsations of the August 29 event as seen by RXTE/PCA ($2-20$ keV). The 
light curve is folded at the period 5.16 s.
The pulse phase spectral analysis results are shown in (b) the OTTB temperature and (c) the total flux. 
Both temperature and flux show significant modulation with pulse phase.
} 
\end{figure*}

\clearpage

\end{document}